\documentclass[11pt]{article}
\usepackage{amsmath}
\usepackage{amsfonts,amssymb,latexsym}

\begin{document}

\newcommand{\nl}{\nonumber\\}
\newcommand{\nnl}{\nl[6mm]}
\newcommand{\nle}{\nl[-2.5mm]\\[-2.5mm]}
\newcommand{\nlb}[1]{\nl[-2.0mm]\label{#1}\\[-2.0mm]}

\renewcommand{\leq}{\leqslant}
\renewcommand{\geq}{\geqslant}

\renewcommand{\theequation}{\thesection.\arabic{equation}}
\let\ssection=\section
\renewcommand{\section}{\setcounter{equation}{0}\ssection}

\newcommand{\be}{\bes}
\newcommand{\ee}{\ees}
\newcommand{\bes}{\begin{eqnarray}}
\newcommand{\ees}{\end{eqnarray}}
\newcommand{\eens}{\nonumber\end{eqnarray}}

\newcommand{\bra}[1]{\big{\langle}#1\big{|}}
\newcommand{\ket}[1]{\big{|}#1\big{\rangle}}
\newcommand{\ave}[1]{\big{\langle}#1\big{\rangle}}
\newcommand{\phys}{\ket{phys}}

\newcommand{\tot}{{Tot}}
\newcommand{\sys}{{Sys}}
\newcommand{\obs}{{Obs}}
\newcommand{\so}{{S-O}}

\renewcommand{\/}{\over}
\renewcommand{\d}{\partial}
\newcommand{\ddt}{{\d\/\d t}}

\newcommand{\no}[1]{{\,:\kern-0.7mm #1\kern-1.2mm:\,}}
\newcommand{\vect}{{\mathfrak{vect}}}
\newcommand{\map}{{\mathfrak{map}}}
\newcommand{\g}{\mathfrak{g}}
\newcommand{\gl}{\mathfrak{gl}}
\newcommand{\Vir}{\mathfrak{Vir}}
\newcommand{\Aff}{\mathfrak{Aff}}

\newcommand{\w} {\omega}
\newcommand{\dlt} {\delta}
\newcommand{\eps} {\epsilon}
\newcommand{\al} {\alpha}
\newcommand{\bt} {\beta}
\newcommand{\gm} {\gamma}
\newcommand{\si} {\sigma}
\newcommand{\mn}{{\mu\nu}}
\newcommand{\ij}{{ij}}

\newcommand{\ab}{{\alpha\beta}}
\newcommand{\cb}{{\gamma\beta}}
\newcommand{\ad}{{\alpha\delta}}
\newcommand{\cd}{{\gamma\delta}}
\newcommand{\ac}{{\alpha\gamma}}
\newcommand{\bd}{{\beta\delta}}

\newcommand{\xx}{{\mathbf x}}
\newcommand{\yy}{{\mathbf y}}
\newcommand{\qq}{{\mathbf q}}
\newcommand{\pp}{{\mathbf p}}
\newcommand{\PP}{{\mathbf P}}
\newcommand{\uu}{{\mathbf u}}
\newcommand{\kk}{{\mathbf k}}
\newcommand{\mm}{{\mathbf m}}
\newcommand{\nn}{{\mathbf n}}
\newcommand{\eell}{{\mathbf l}}

\newcommand{\dx}{d^{d+1}\!x}
\newcommand{\dy}{d^{d+1}\!y}
\newcommand{\dk}{d^{d+1}\!k}
\newcommand{\dxx}{d^d\!x}
\newcommand{\dyy}{d^d\!y}
\newcommand{\dkk}{d^d\!k}

\newcommand{\e}{{\mathrm e}}
\newcommand{\ext}{{\mathrm{\,ext}}}
\newcommand{\tr}{{\mathrm {tr}\,}}
\newcommand{\half}{{1\/2}}
\newcommand{\quart}{{1\/4}}
\newcommand{\stwo}{{1\/\sqrt2}}

\newcommand{\noll}{{\hat 0}}
\newcommand{\zero}{{\mathbf 0}}

\renewcommand{\L}{{\mathcal L}}
\newcommand{\J}{{\mathcal J}}
\newcommand{\EE}{{\mathcal E}}

\newcommand{\fa}{\phi_\alpha}
\newcommand{\fb}{\phi_\beta}
\newcommand{\pa}{\pi_\alpha}
\newcommand{\pb}{\pi_\beta}

\newcommand{\tphi}{\tilde\phi}
\newcommand{\tpsi}{\tilde\psi}
\newcommand{\tpi}{\tilde\pi}
\newcommand{\tchi}{\tilde\chi}

\newcommand{\stwk}{\sqrt{2|\kk|}}

\newcommand{\fsa}{\phi^*_\alpha}
\newcommand{\psb}{\pi_*^\beta}

\newcommand{\RR}{{\mathbb R}}
\newcommand{\CC}{{\mathbb C}}
\newcommand{\ZZ}{{\mathbb Z}}
\newcommand{\NN}{{\mathbb N}}

\title{{The physical observer II: Gauge and diff anomalies}}

\author{T. A. Larsson \\
Vanadisv\"agen 29, S-113 23 Stockholm, Sweden\\
email: thomas.larsson@hdd.se}

\maketitle
\begin{abstract}
In a companion paper we studied field theory in the presence of a
physical observer with quantum dynamics. Here we describe the most
striking consequence of this assumption: new gauge and diff anomalies
arise. The relevant cocycles depend on the observer's spacetime
trajectory and can hence not appear in QFT, where this quantity is never
introduced. Diff anomalies necessarily arise in every locally
nontrivial, non-holographic theory of quantum gravity. Cancellation of
the divergent parts of the anomalies only works if spacetime has four
dimensions.
\end{abstract}

\vskip 3cm

\section{Introduction}

In a companion paper \cite{Lar08a},	the notion of absolute and relative
fields was introduced. QFT deals with absolute fields $\phi_A(t,\xx)$,
where the location $\xx$ is measured relative to a fixed 
origin\footnote{A fixed origin may be regarded as the location of an
infinitely massive observer. This is a hidden assumption about an
infinite observer mass in QFT.}, using some measuring rods. In contrast, 
QJT (Quantum Jet Theory) deals with relative fields $\phi_R(t,\xx)$, 
labelled by a location measured relative to a physical observer's 
position $\qq(t)$.  A physical observer obeys some
quantum dynamics, and hence its position at a given time is a complete
observable, which can be predicted by the theory and which becomes an
operator after quantization. The difference between absolute and relative
fields hence resides ``at the other side of the measuring rod''; this
is a fixed origin in QFT but a quantized observable in QJT.

Matrix elements in QJT depend on the observer's physical properties, in
particular on its mass $M$ and charge $e$. These properties are never
mentioned in QFT, which means that some tacit assumption is made; QFT is
recovered from QJT in the joint limit $M \to \infty$ and $e \to 0$. This
limit is well defined for all interactions except gravity, where mass and
charge are related; inert mass equals heavy mass. Hence the QFT limit
of QJT does not exist specifically in the presence of gravity. This is
the origin of the difficulties with applying QFT to gravity.
 
The most striking new feature in QJT is the appearence of new gauge and
diff anomalies, which have no counterpart in QFT. In all known
representations of the extended gauge and diffeomorphism algebras, the
relevant cocycles\footnote{We use the terms ``cocycle'', ``anomaly'' and
``extension'' interchangably.} are functionals of the observer's
trajectory in spacetime. They can not be formulated within a QFT
framework, since the observer is never introduced in QFT, but they arise
naturally in QJT. The presence of new anomalies proves that QJT is
substantially different from QFT.

The new anomalies only appear if we consider gauge transformations or
diffeomorphisms in spacetime; the spatial subalgebras are essentially
anomaly free.
Constraint algebras in canonical quantization on a fixed foliation do
hence not see these anomalies. However, it is possible to recover the
gauge anomalies in QJT by moving away infinitesimally form the 
equal-time surface. This point-splitting construction is the main 
result in this paper.

We end this paper with a discussion on gauge anomalies and consistency,
which contrary to popular belief are not mutually exclusive.

\section{ Gauss' law }

\subsection{ Free electromagnetic field}

In \cite{Lar08a} we quantized the free electromagnetic field within QJT
by fixing a gauge. However, it is often more convenient to quantize 
first and impose the constraints afterwards. As a warmup, we review
how this is done for the free electromagnetic field, within QFT rather
than QJT. All fields are hence absolute fields.

The canonically conjugates are the gauge potential $A_i(\xx)$ and the
electric field $E_i(\xx)$, with nonzero commutators
\be
[A_i(\xx), E_j(\yy)] = i\dlt_\ij \dlt(\xx-\yy).
\ee
The Hamiltonian reads
\be
H = \int \dxx\, \big( \half E_i(\xx) E_i(\xx) 
+ \quart F_\ij(\xx) F_\ij(\xx) \big),
\ee
where $F_\ij = \d_i A_j - \d_j A_i$. The fields are not independent, but
subject to the Gauss' law constraint
\be
J(\xx) \equiv \d_i E_i(\xx) \approx 0.
\ee
We quantize the theory by replacing Poisson brackets by commutators, 
passing to Fourier space and demanding that negative-frequency modes 
annihilate the vacuum. If we introduce the magnetic field
$B_i = \eps_{ijk} F_{jk}$, the Hamiltonian becomes
\be
H = \half \int \dkk\, \bigg( E_i(\kk) E_i(-\kk) 
+  B_i(\kk) B_i(-\kk) \bigg),
\ee
and 
\be
[E_i(\kk), B_j(\kk')] = \eps_{ijm} k_m \dlt(\kk+\kk').
\ee
These brackets are compatible with Gauss' law in its Fourier form:
\be
J(\kk) = k_i E_i(\kk) \approx 0.
\label{GaussEM}
\ee
We now introduce the oscillators
\bes
a_i(\kk) &=& {1\/\stwk}(E_i(\kk) - i |\kk| A_i(\kk)), 
\nlb{aaEM}
a^\dagger_i(\kk) &=& {1\/\stwk}(E_i(\kk) + i |\kk| A_i(\kk)),
\eens
with commutators
\be
[a_i(\kk), a^\dagger_j(\kk')] =\dlt_\ij(\kk)\dlt(\kk+\kk').
\ee
The normal-ordered Hamiltonian becomes a sum of noninteracting 
harmonic oscillators,
\be
H = \int \dkk\, |\kk| a^\dagger_i(\kk) a_i(-\kk).
\ee
We posit that the vacuum $\ket0$ is annihilated by all negative
frequency states, i.e. $a_i(\kk)\ket0 = 0$. Unlike the oscillators
constructed in our companion paper \cite{Lar08a}, the oscillators
(\ref{aaEM}) are not immediately compatible with Gauss' law 
(\ref{GaussEM}), which takes the form
\be
J(\kk) = \sqrt{|\kk|\/2} ( k_i a_i(\kk) + k_i a^\dagger_i(\kk) ).
\label{JkEM}
\ee
Instead of realizing the constraint $J(\kk) = 0$ as an operator
equation, we impose it as a condition on physical states; by definition,
a state $\phys$ is physical if it satisfies $J(\kk)\phys = 0$, and
two physical states are equivalent if they differ by a state of the
form $J(\kk)\ket{}$. For this definition to be self-consistent, the
constraint must commute with the Hamiltonian and itself:
\be
[J(\kk), J(\kk')] = [J(\kk), H] = 0.
\label{JJEM}
\ee
It is readily verified that these relations continue to hold after
quantization. It is clear that quantization
can not destroy the validity of (\ref{JJEM}) because the Gauss law 
generators (\ref{JkEM}) are linear in the oscillators, so there is no
need for normal ordering. In interacting theories the constraint 
generators are at least bilinear in oscillators, and normal ordering 
can potentially invalidate the analogue of (\ref{JJEM}).

\subsection{ Yang-Mills field }

Let us generalize this well-known story to Yang-Mills theory based on a
finite-dimensional Lie algebra $\g$. We denote the generators by $J^a$
and structure constants by $f^{abc}$, and we assume that $\g$ has a 
Killing metric $\dlt^{ab}$. The Lie brackets are thus
\be
[J^a, J^b] = if^{abc} J^c.
\label{g}
\ee
From our point of view, the important new feature is that the Gauss 
law constraint also contains a bilinear term:
\be
J^a(\xx) \equiv \d_i E^a_i(\xx) + f^{abc} A^b_i(\xx) E^c_i(\xx),
\label{GaussYM}
\ee
where the nonzero CCR read
\be
[A^a_i(\xx), E^b_j(\yy)] = i\dlt^{ab} \dlt_\ij \dlt(\xx-\yy).
\ee
The constraints (\ref{GaussYM}) satisfy the current algebra $\map(d,\g)$
(algebra of maps from $d$-dimensional space to $\g$):
\be
[J^a(\xx), J^b(\yy)] = if^{abc} J^c(\xx) \dlt(\xx-\yy).
\label{mapg}
\ee
We again pass to Fourier space, where CCR become
\be
[A^a_i(\kk), E^b_j(\kk')] 
&=& i\dlt^{ab} \dlt_\ij \dlt(\kk+\kk').
\ee
We now introduce the oscillators
\bes
a^a_i(\kk) &=& {1\/\stwk}(E^a_i(\kk) - i |\kk|A^a_i(\kk)), \nle
a^{\dagger a}_i(\kk) &=& {1\/\stwk}(E^a_i(\kk) + i |\kk|A^a_i(\kk)),
\eens
with commutators
\be
[a^a_i(\kk), a^{\dagger b}_j(\kk')] = \dlt^{ab} \dlt_\ij \dlt(\kk+\kk').
\ee
By definition, the vacuum $\ket0$ is annihilated by all negative
frequency states, i.e. $a_i(\kk)\ket0 = 0$. However, now we encounter 
a problem with the second term of (\ref{GaussYM}), which after normal
ordering reads in Fourier space
\bes
J^a(\kk) &=& f^{abc}\int \dkk'\, \no{A^b_i(\kk') E^c_i(\kk-\kk')} 
\label{JaYM}\\
&=& if^{abc}\int \dkk'\, {1\/2}\big( 
a^b_i(\kk') a^c_i(\kk-\kk') 
- a^{\dagger b}_i(\kk') a^c_i(\kk-\kk') +\nl
&& +\ a^{\dagger c}_i(\kk-\kk') a^b_i(\kk') 
- a^{\dagger b}_i(\kk') a^{\dagger c}_i(\kk-\kk')	\big).
\eens
These generators satisfy the gauge algebra with two normal-ordering
contributions,
\be
[J^a(\kk), J^b(\eell)] = if^{abc} J^c(\kk+\eell) + \ext_1 + \ext_2
\label{JaJb}
\ee
where
\be
\ext_1 = -\ext_2 = Q \dlt^{ab} \dlt(\kk+\eell) \int \dkk'\, 1,
\ee
and $Q$ denotes the second Casimir operator in the adjoint 
representation: $f^{acd}f^{bcd} = Q\dlt^{ab}$. Since the two extensions
in (\ref{JaJb}) cancel, there is no anomaly. However, care must be
taken, because both terms are proportional to $\int \dkk'\,1 = \infty$,
so $\ext_1 + \ext_2$ is a constant of the form $\infty - \infty$. 
We will explain in section \ref{sec:split} below how to turn this
difference into a finite term within the framework of QJT.

That the total extension vanishes is of course not surprising, because
the first and last term in (\ref{JaYM}) vanish. E.g.,
$a^b_i(\kk') a^c_i(\kk-\kk')$ is symmetric under the replacement
$\kk' \to \kk-\kk'$, $b \leftrightarrow c$, and yields zero when 
multiplied with	the antisymmetric constant $f^{abc}$. The strategy
for constructing a nonzero extension therefore consists of avoiding
this cancellation.

\section{ General bilinear gauge generators }
\label{sec:general}

The need for normal ordering is not unique to Yang-Mills theory. In fact,
it is a generic feature of all interacting theories, whenever the
constraint generators are at least bilinear in the oscillators. Only
for the free electromagnetic field, where the constraint is linear in
$E_i$, is normal ordering unnecessary, and the classical constraint
runs no risk of breaking down upon quantization. To study gauge anomalies,
we must hence turn to interacting theories. 

Consider a set of fields $\fa$ with canonical conjugate momenta $\pb$.
The CCR read ($\hbar=1$ throughout this paper)
\be
[\fa, \pb] = i\dlt_\ab, \qquad [\fa,\fb] = [\pa,\pb] = 0.
\ee
We use an abbreviated notation, where indices $\al$, $\bt$ are shorthand
for both discrete and continuous indices; in particular, this includes
the space coordinates. When we want to emphasize that some of the indices
are continuous, we can always make the substitutions $\fa \to \fa(\xx)$, 
$\pb \to \pb(\yy)$, etc., and remember that contraction also implies
integration over continuous coordinates. The main difference between 
discrete and continuous is that the product of delta functions,
\be
\dlt_\ab \dlt_{\bt\al} \to \dlt(\xx-\yy)\dlt(\yy-\xx)
 = \dlt(\zero)\dlt(\xx-\yy), \qquad \hbox{(no sum on $\al,\bt$)}
\label{double}
\ee
is proportional to $\dlt(\zero)$ and hence ill defined in the continuous
case. Denote by $N$ the number of degrees of freedom
that the index $\al$ runs over; if $\al$ also includes continuous
degrees of freedom, $N = \infty$.

Assume that our constraint algebra takes the form
\be
[J^a, J^b] = if^{abc}J^c.
\label{alg}
\ee
This is formally of the form (\ref{g}), but
in view of our abbreviated notation it is a shorthand for the current
algebra (\ref{mapg}). For simplicity, we only consider the case that
the constraint algebra is a proper Lie algebra. This evidently includes
Yang-Mills theory, but also the constraint algebra of general relativity 
can be cast in Lie-algebraic form \cite{KR95,Mar96}.
Assume that the matrices $M^a = (M^a_\ab)$ furnish a representation of 
our constraint algebra, i.e.
\be
[M^a, M^b]_\ab = M^a_{\al\gm} M^b_{\gm\bt} - M^a_{\al\gm} M^b_{\gm\bt}
= if^{abc} M^c_\ab.
\label{MM}
\ee
Then the operators
\be
J^a = M^a_\ab E_\ab \equiv i M^a_\ab \fa\pb
\ee
satisfy the algebra (\ref{alg}). We have introduced the bilinear
combinations 
\be
E_\ab = i\fa\pb,
\label{Eab}
\ee
which satisfy the algebra $\gl(N)$:
\be
[E_\ab, E_\cd] = \dlt_\cb E_\ad - \dlt_\ad E_\cb.
\label{glN}
\ee
The field operators carry a representation of $\gl(N)$:
\bes
[E_\ab, \phi_\gm] &=& \dlt_\cb \fa, \nle
[E_\ab, \pi_\gm] &=& -\dlt_{\al\gm}\pb,
\eens
as well as a representation of $\g$:
\bes
[J^a, \fa] &=& M^a_\ab \fb = (M^a\phi)_\al, \nle
{[}J^a, \pa] &=& - M^a_{\bt\al} \pb = -(\pi M^a)_\al.
\eens
Introduce the oscillators
\be
a_\al = {1\/\sqrt2} (\fa + i\pa), \quad
a^\dagger_\al = {1\/\sqrt2} (\fa - i\pa), 
\ee
so that
\be
\fa = {1\/\sqrt2}(a_\al + a^\dagger_\al), \quad
\pa = -{i\/\sqrt2}(a_\al - a^\dagger_\al).
\ee
Upon quantization we must normal order, and move the creation operators
$a^\dagger_\al$ to the left of the annihilation operators $a_\al$. The
$\gl(N)$ operators (\ref{Eab}) are replaced by
\be
E_\ab &=& i\no{\fa\pb} 
= \half(a_\al a_\bt + a^\dagger_\al a_\bt 
- a^\dagger_\bt a_\al - a^\dagger_\al a^\dagger_\bt ).
\label{:Eab:}
\ee
This amounts to adding a constant to $E_\ab$:
\be
E_\ab \to E_\ab + [a_\al, a^\dagger_\bt] = E_\ab + \dlt_\ab.
\label{Eab-add}
\ee
The bracket (\ref{glN}) receives two new contributions due to 
normal ordering:
\bes
[a_\al a_\bt, a^\dagger_\gm a^\dagger_\dlt] &=&
(\dlt_\cb\dlt_\ad + \dlt_\bd\dlt_\ac) + ...,
\nlb{nocontrib}
[a^\dagger_\al a^\dagger_\bt, a_\gm a_\dlt] &=&
-(\dlt_\cb\dlt_\ad + \dlt_\bd\dlt_\ac) + ...,
\eens
where ellipses denote terms that are already present before
normal ordering. The sum of these two contributions vanishes, and the
normal-ordered operators (\ref{:Eab:}) satisfy $\gl(N)$ without any
extra terms. This is also clear from (\ref{Eab-add}).

However, we must be careful when we deal with infinitely many
degrees of freedom, $N = \infty$. In particular, the bracket 
$[E_{\al\al}, E_{\bt\bt}]$ gets two contributions (\ref{nocontrib})
that are proportional to the product of delta-functions (\ref{double}), 
and this signals a problem in the continuous case.
If we restore the space coordinates, the $\gl(\infty)$ generators become
\be
E_\ab(\xx,\yy) &=& i\no{\fa(\xx)\pb(\yy)},
\ee
and the current algebra generators are
\be
J^a(\xx) = M^a_\ab E_\ab(\xx,\xx)=
\int \dxx'\, M^a_\ab E_\ab(\xx,\xx')\dlt(\xx-\xx')
\label{Jphi}
\ee
The normal ordering contributions (\ref{nocontrib}) become
\bes
[a_\al(\xx) a_\bt(\xx), a^\dagger_\gm(\yy) a^\dagger_\dlt(\yy)] &=&
(\dlt_\cb\dlt_\ad + \dlt_\bd\dlt_\ac) \dlt(\xx-\yy)\dlt(\xx-\yy) + ...,
\nlb{nlfield}
[a^\dagger_\al(\xx) a^\dagger_\bt(\xx), a_\gm(\yy) a_\dlt(\yy)] &=&
-(\dlt_\cb\dlt_\ad + \dlt_\bd\dlt_\ac)\dlt(\xx-\yy)\dlt(\xx-\yy)  + ...,
\eens
The important observation is that both terms are proportional to
$\dlt(\xx-\yy)\dlt(\xx-\yy) = \dlt(\zero)\dlt(\xx-\yy)$. Although
the sum of the two terms in (\ref{nlfield}) vanishes, each term is 
proportional to $\dlt(\zero)$ and thus infinite. Again, this is a
signal that care is needed.

\section{ Extensions of gauge algebras }
\label{sec:ext} 

It is in a sense surprising that the generators (\ref{Jphi}) satisfy the
current algebra without anomalous terms, because mathematically such terms
do exist, and we expect that anything that can happen will happen in 
quantum theory. It is well known and easy to
verify that the spacetime version of (\ref{mapg}) admits a central 
extension \cite{EF94,LMNS1,LMNS2,PS86}:
\bes
[J^a(t,\xx), J^b(t',\xx')] &=& if^{abc} J^c(t,\xx) \dlt(t-t')\dlt(\xx-\xx')
\nlb{KMd1}
&&+\ K \dlt^{ab} \dot\dlt(t-t')\dlt(\xx-\xx').
\eens
Unlike the situtation in one dimension, the extension is no longer central 
when we take Poincar\'e or diffeomorphism symmetry into account, because
the ``central'' term does not commute with spacetime transformations. 
However, (\ref{KMd1}) admits a covariant formulation, which can be 
written in Fourier space as
\bes
[J^a(k), J^b(\ell)] &=& if^{abc} J^c(k+\ell) 
- K\dlt^{ab} k_\mu S^\mu(k+\ell), \nl
{[}J^a(k), S^\nu(\ell)] &=& {[}S^\mu(k), S^\nu(\ell)]\ =\ 0, 
\label{KMd}\\
k_\mu S^\mu(k) &\equiv& 0.
\eens
Here $k = (k_\mu) \in \ZZ^{d+1}$ labels the Fourier modes on a 
$(d+1)$-dimensional torus; the constant is denoted by
a capital $K$ to avoid confusion with Fourier
labels. Since (\ref{KMd}) is a generalization of affine
Kac-Moody algebras to $(d+1)$ dimensions, we denote it by $\Aff(d+1,\g)$;
the usual affine algebra is $\hat \g = \Aff(1,\g)$.
To recover the delta-function form from this extension, assume that
$S^\mu(k)$ is of the form
\be
S^\mu(k) = \dlt^\mu_0 \dlt(k_0) S(\kk).
\label{Sspat}
\ee
This expression clearly satisfies $k_\mu S^\mu(k) \equiv 0$, and the 
extension in the $JJ$ bracket takes the form 
$-K \dlt^\ab k_0 \delta(k_0+\ell_0) S(\kk+\eell)$, which is the
Fourier transform of (\ref{KMd1}) provided that we choose 
$S(\kk) = \dlt(\kk)$.

In the formulation (\ref{KMd}), $\Aff(d+1,\g)$
admits an intertwining action of diffeomorphisms. Denote the $\vect(d+1)$
(algebra of vector fields on the $(d+1)$-dimensional torus) generators 
by $L_\mu(m) = -i\exp(im\cdot x)\d_\mu$. Its semi-direct product with
the current algebra (\ref{KMd}) is defined by the brackets
\bes
[L_\mu(k), L_\nu(\ell)] &=& \ell_\mu L_\nu(k+\ell) - k_\nu L_\mu(k+\ell) \nl 
&& +\ (c_1 k_\nu \ell_\mu + c_2 k_\mu \ell_\nu) k_\rho S^\rho(k+\ell), \nl
{[}L_\mu(k), S^\nu(\ell)] &=& \ell_\mu S^\nu(k+\ell)
 + \delta^\nu_\mu k_\rho S^\rho(k+\ell), 
\label{mVir}\\
{[}L_\mu(k), J^a(\ell)] &=& \ell_\mu J^a(k+\ell).
\eens
Note that we have included two abelian extensions, which makes this 
algebra a multi-dimensional generalization of the Virasoro algebra, 
denoted by $\Vir(d+1)$
\cite{Lar91,RM94}. It is straightforward to verify that (\ref{KMd}) and
(\ref{mVir}) satisfy the axioms for a Lie algebra, and that in the 
one-dimensional case both extensions in (\ref{mVir}) reduce to the usual 
Virasoro algebra:
\be
[L_k, L_\ell] = (\ell-k) L_{k+\ell} - {c\/12} (k^3-k) \delta_{k+\ell},
\label{Vir}
\ee
apart from the trivial, linear cocycle which can be absorbed into a
redefinition of $L_0$.

Since $S^\mu(m)$ does not commute with diffeomorphisms, the condition
(\ref{Sspat}) is not compatible with the full algebra $\Vir(d+1)$. 
However, it is preserved by the spatial subalgebra $\vect(d)$ with 
generators $L_i(\kk)$, where the time component $k_0 = 0$; $k = (0,\kk)$. 
The condition (\ref{Sspat}) is equivalent to demanding that 
$S^j(\ell) = 0$ for all $\ell$ including $\ell_0 \neq 0$. This is 
consistent because
\be
[L_i(\kk), S^j(\ell)] = \ell_i S^j(\kk+\ell) 
+ \delta^j_i\big( \ell_0 S^0 (\kk+\ell) + \ell_n S^n(\kk+\ell) \big),
\ee
and the RHS vanishes since $S^0 (\kk+\ell) \propto \delta(\ell_0)$
and $S^j(\kk+\ell) = 0$ by assumption.
The time component $S^0(\ell) = 0$ unless
$\ell_0 = 0$, and $S(\eell)$ in (\ref{Sspat}) transforms as a scalar
density under spatial diffeomorphisms. The full anomalous algebra of
spacetime gauge transformations and spatial diffeomorphisms becomes
\bes
[J^a(k), J^b(\ell)] &=& if^{abc} J^c(k+\ell) 
- K \dlt^{ab} k_0 \dlt(k_0+\ell_0) S(\kk+\eell), \nl
{[}L_i(\kk), J^a(\ell)] &=& \ell_i J^a(\kk+\ell), \nl
{[}L_i(\kk), S(\eell)] &=& \ell_i S(\kk+\eell), 
\label{spat}\\
{[}L_i(\kk), L_j(\eell)] &=& 
 \ell_i L_j(\kk+\eell) - k_j L_i(\kk+\eell), \nl
{[}J^a(k), S(\eell)] &=& {[}S(\kk), S(\eell)] = 0.
\eens
We can here consistently assume that $S(\kk) = \dlt(\kk)$; this is
the largest subalgebra of $\Vir(d+1)\ltimes\Aff(d+1,\g)$ for which the
extension is central. In the subalgebra of spatial gauge transformations,
generated by $J^a(\kk)$ with $k_0 = 0$, the extension disappears 
completely. This leads to the important conclusion that the new gauge
anomalies only arise if we consider spacetime transformations;
in a purely spatial constraint algebra, which arises if we only consider
fields on a fixed foliation, the new gauge anomalies vanish. The
conclusion is morally the same for the new diff anomalies, except that
working on a fixed foliation does not make much sense in this case, 
since spacetime diffeomorphisms do not preserve the foliation. A better
alternative would be to work with a covariant formalism. Indeed, this
was the main motivation for MCCQ, cf. subsection \ref{ssec:MCCQ}.

It must be emphasized that these extensions are not equivalent to the
usual types of gauge anomalies arising in QFT. In particular, the Virasoro
extension (\ref{mVir}) is defined in any number of dimensions, but in QFT
there are no diff anomalies at all in four dimensions \cite{Bon86}.
Moreover, gauge anomalies in Yang-Mills theory are proportional to the
third Casimir $d^{abc} = \tr \{J^a, J^b\} J^c$ \cite{Wein96}, whereas the
$\Aff(d+1,\g)$ extension is proportional to the second Casimir $\dlt^\ab
\propto \tr J^aJ^b$. In the Hamiltonian formalism, conventional gauge
anomalies in three dimensions give rise to a Mickelsson-Faddeev (MF) 
algebra \cite{Mi89}, which can be written in Fourier space as
\bes
[J^a(\kk), J^b(\eell)] &=& if^{abc} J^c(\kk+\eell) 
+ d^{abc} \eps^{ijn} k_i \ell_j A^c_n(\kk+\eell), \nl
{[}J^a(\kk), A^b_i(\eell)] &=& if^{abc} A^c_i(\kk+\eell)
+ \dlt^{ab} k_i \dlt(\kk+\eell), \nl
{[}A^a_i(\kk), A^b_j(\eell)] &=& 0.
\label{MF}
\ees
Here $A^a_i(\kk)$ denotes the Fourier modes of the gauge connection. It 
is clear that the extensions (\ref{KMd}) and (\ref{MF}) are essentially
different. Gauge anomalies of the MF form render
the theory inconsistent and must be avoided. There is a simple
mathematical reason for this: the MF algebra does not admit any nontrivial
unitary representations on a separable Hilbert space \cite{Pic89}, and
hence it can not be a symmetry of a quantum theory. This no-go theorem
does not apply to the substantially different algebras (\ref{KMd}) and
(\ref{mVir}).

Why can the extensions in (\ref{KMd}) and (\ref{mVir}) only arise within
QJT and not within QFT? In all known representations, the anomaly takes the 
form
\be
S^\mu(k) = \int dt\ \dot q^\mu(t) \exp(ik\cdot q(t)),
\ee
where $q^\mu(t)$ denotes the observer's trajectory in spacetime
\cite{Lar98}. This can only be written down if the observer's position
has been introduced in the first place, i.e. if we pass from QFT to QJT.
In this realization, the last condition in (\ref{KMd}) corresponds to
Stokes' theorem.

\section{ Origin of gauge anomalies }

We learned in the previous section that the algebra of gauge 
transformations admits extensions, but that these are not realized
in QFT, because the two anomalous terms in (\ref{JaJb}) cancel. It is
natural to ask if there is some way to avoid this conclusion and 
realize the anomalous terms. Indeed there is, and the crucial idea 
is chirality in a general sense. By starting from twice as many 
oscillators, but only acting with our symmetry algebra on half of them,
we can avoid anomaly cancellation. Normal ordering in a chiral theory
only gives rise one of the contributions in (\ref{nocontrib}) or 
(\ref{nlfield}), and no cancellation occurs.

After making the general construction in this section, both for bosonic
and fermionic degrees of freedom, we apply the idea
to genuinely chiral theories in section \ref{sec:chiral}. In section
\ref{sec:split} we show how a very similar idea yields a central
extension of the gauge algebra of the form (\ref{spat}). This 
construction, which only works in QJT, amounts to moving away
an infinitesimal distance from the equal-time surface.

\subsection{ Bosons}

We posit four types of oscillators $a_\al, a^\dagger_\al, b_\al,
b^\dagger_\al$, subject to the commutators
\bes
[a_\al, b^\dagger_\bt] &=& c_\ab, \nl
{[}a^\dagger_\al, b_\bt] &=& - c^*_\ab, \nl
{[}a_\al, a^\dagger_\bt] &=& [b_\al, b^\dagger_\bt] 
\ =\ [a_\al, b_\bt]\ =\ [a^\dagger_\al, b^\dagger_\bt]\ =\ 0, 
\label{chirality}\\
{[}a_\al, a_\bt] &=& [b_\al, b_\bt] 
\ =\ [a^\dagger_\al, a^\dagger_\bt]
\ =\ [b^\dagger_\al, b^\dagger_\bt]\ =\ 0.
\eens
Here $c_\ab$ are some complex constants and $c^*_\ab$ their complex 
conjugates. As the notation suggests, we assume that hermitean 
conjugation acts as $a_\al \to a^\dagger_\al$, $b_\al \to b^\dagger_\al$. 
Further assume that the structure constants have the form
\be
c_\ab = \dlt_\ab + i\gm_\ab, \qquad 
c^*_\ab = \dlt_\ab - i\gm_\ab,
\label{cc}
\ee
where $\dlt_\ab$ denotes the Kronecker delta, and both $\dlt_\ab$ and
$\gm_\ab$ are real. The real combinations
\bes
\fa &=& \phi^\dagger_\al = {1\/\sqrt2}(a_\al + a^\dagger_\al), 
\nle
\pa &=& \pi^\dagger_\al = -{i\/\sqrt2}(b_\al - b^\dagger_\al),
\eens
satisfy the usual CCR
\bes
[\fa, \pb] &=& {i\/2}(c_\ab + c^*_\ab) = i\dlt_\ab, \nle
{[}\fa, \fb] &=& [\pa,\pb] = 0.
\eens
The normal-ordered combinations
\be
E_\ab &=& i\no{\fa\pb} 
= \half(a_\al b_\bt + a^\dagger_\al b_\bt 
- b^\dagger_\bt a_\al - a^\dagger_\al b^\dagger_\bt ).
\label{noEab}
\ee
therefore satisfies some central extension of $\gl(N)$ (\ref{glN}). The
extension is necessarily central since $E_\ab$ is bilinear in the 
oscillators. There are two normal-ordering contributions to the extension:
\bes
[a_\al b_\bt, a^\dagger_\gm b^\dagger_\dlt] &=& c^*_\cb c_\ad + ..., \nle
{[}a^\dagger_\al b^\dagger_\bt, a_\gm b_\dlt] &=& - c_\cb c^*_\ad + ....
\eens
The normal-ordered generators (\ref{noEab}) hence satisfy the following
central extension of $\gl(N)$
\be
[E_\ab, E_\cd] = \dlt_\cb E_\ad - \dlt_\ad E_\cb 
+ \quart(c^*_\ad c_\cb - c^*_\cb c_\ad).
\ee
In view of (\ref{cc}), the extension can alternatively be written as
\be
{i\/2}( \dlt_\ad \gm_\cb - \dlt_\cb \gm_\ad).
\label{bosext}
\ee
In this form, it is clear that the extension is real and nonzero. 
Normal ordering corresponds to the replacement
\be
E_\ab \to E_\ab + \half[a_\al, b^\dagger_\bt] = E_\ab + \half c_\ab.
\ee 
If the index $\al$ only runs over a finite set of values, such a 
redefinition can of course not result in anything non-trivial.
However, we will see that this results in a non-trivial current
algebra extension in the limit of infinitely many degrees of freedom.

What gives us a central extension in this case is a kind of chirality.
We have four oscillators $a_\al, a^\dagger_\al, b_\al, b^\dagger_\al$ 
but only two fields $\fa$ and $\pa$. Hence it must be possible to define
two more fields, which are inert under $\gl(N)$. Indeed, the full set of
real fields that we can write down is
\bes
\fa = {1\/\sqrt2}(a_\al + a^\dagger_\al), 
&\qquad&
\pa = -{i\/\sqrt2}(b_\al - b^\dagger_\al),
\nle
\psi_\al = {i\/\sqrt2}(a_\al - a^\dagger_\al), 
&\qquad&
\chi_\al = {1\/\sqrt2}(b_\al + b^\dagger_\al),
\eens
These fields satisfy the algebra
\bes
[\fa, \pb] = i\dlt_\ab, 
&\qquad&
[\fa, \chi_\bt] = i\gm_\ab
\nlb{nondiag}
{[}\psi_\al, \pb] = -i\gm_\ab, 
&&
{[}\psi_\al, \chi_\bt] = i\dlt_\ab.
\eens
It is not quite obvious that this is a Heisenberg algebra, due to the
extra terms proportional to $\gm_\ab$ in the RHS. However, the equations
(\ref{nondiag}) can be diagonalized, showing that they do indeed describe
two independent canonically conjugate pairs. To this end, it is convenient
to go over to index-free notation, and introduce the vectors 
$\phi = (\fa)$, etc. The algebra (\ref{nondiag}) then becomes
\bes
[\phi,\pi] = i, \qquad
[\phi,\chi] = i \gm, 
\nlb{nondiag2}
{[}\psi,\pi] = -i\gm, \qquad
[\psi,\chi] = i.
\eens
To diagonalize these relations, we introduce two canonically conjugate
pairs $\tphi, \tpi$ and $\tpsi, \tchi$, whose only nonzero brackets are
$[\tphi,\tpi] = [\tpsi,\tchi] = i$. Let
\bes
\phi = \stwo(\tphi + \si_1\tpsi), &\qquad&
\pi = \stwo(\tpi + \tchi\si_1), 
\nlb{diag}
\psi = \stwo(i A\si_2\tpsi + A\si_3\tphi) &&
\chi = \stwo(-i\tchi\si_2 \bar A + \tpi\si_3\bar A).
\eens
where $\si_i$ are the Pauli matrices satisfying 
$\si_i\si_j = \dlt_\ij + i\eps_{ijk}$, and $A$ is a unitary matrix
anticommuting with $\si_3$, i.e. $A\bar A = \bar A A = 1$ and
$A\si_3 + \si_3 A = 0$.
It is straightforward to verify that the combinations (\ref{diag}) 
indeed satisfy the relations (\ref{nondiag2}).
An explicit represention in terms of $4\times4$ Dirac matrices is 
given by
\be
1 = \begin{pmatrix} 1 & 0 \\ 0 & 1 \end{pmatrix}, 
\ 
\si_i = \begin{pmatrix} \si_i & 0 \\ 0 & \si_i \end{pmatrix}, 
\ 
A = \begin{pmatrix} 0 & 1 \\ -1 & 0 \end{pmatrix}, 
\ 
\bar A = \begin{pmatrix} 0 & -1 \\ 1 & 0 \end{pmatrix},
\ee 
where each block is a $2\times2$ matrix and $\si_i$ denotes the Pauli
matrices. $A = \gm^0\gm^5$ in the Dirac representation of the Dirac
matrices.

\subsection{Fermions}

The analysis above readily carries over to fermionic oscillators with
some sign changes. For brevity, we just list the results in index-free
notation. We introduce oscillators $a, a^\dagger, b, b^\dagger$ with 
nonzero canonical anticommutation relations (CAR)
\be
\{a,b^\dagger\} = c, \qquad 
\{a^\dagger, b\} = \{b, a^\dagger\} = c^\dagger, 
\ee
where $c = 1+i\gm$ and $c^\dagger = 1-i\gm$. From these oscillators we
can construct the four fields
\bes
\phi = {1\/\sqrt2}(a+a^\dagger), &\qquad&
\pi = {1\/\sqrt2}(b+b^\dagger), \nle
\psi = {i\/\sqrt2}(a-a^\dagger), &\qquad&
\chi = {i\/\sqrt2}(b-b^\dagger).
\eens
which have the nonzero brackets
\bes
\{\phi,\pi\} = 1, &\qquad&
\{\phi,\chi\} = \gm, \nle
\{\psi,\pi\} = -\gm, &\qquad&
\{\psi,\chi\} = 1.
\eens
To diagonalize these relations, we introduce two canonically conjugate
pairs $\tphi, \tpi$ and $\tpsi, \tchi$, whose only nonzero brackets are
$\{\tphi,\tpi\} = \{\tpsi,\tchi\} = 1$. The relation to the 
original fields is
\bes
\phi = \stwo(\tphi + \si_1\tpsi), &\qquad&
\pi = \stwo(\tpi + \tchi\si_1), \nle
\psi = \stwo(i A\si_2\tpsi + A\si_3\tphi) &&
\chi = \stwo(-i\tchi\si_2 \bar A + \tpi\si_3\bar A).
\eens
where $\si_i$ and $A$ are the same as in the bosonic case.

The $\gl(N)$ generators read
\be
E = \no{\phi\otimes\pi}
= \half(a \otimes b + a^\dagger\otimes b 
-(1\otimes b^\dagger)(a\otimes 1) + a^\dagger\otimes b^\dagger).
\ee
These operators satisfy the algebra $\gl(N)$ (\ref{glN}), and the extension
\be
\quart(c^*_\cb c_\ad - c^*_\ad c_\cb) = 
{i\/2}( \dlt_\cb \gm_\ad - \dlt_\ad \gm_\cb)
\ee
is the negative of the corresponding bosonic extension. This is a
generic feature: if we interchange bosons and fermions, anomalies change
sign but retain amplitudes.

\section{ Chiral theories }
\label{sec:chiral}

If the indices $\al,\bt$ belong to some finite set, the central extensions
in the previous section can be removed by a redefinition and are hence
trivial. However, we are interested in the case that the index set is
infinite, and includes the space coordinates. Therefore, we briefly repeat
the analysis for bosons with the spatial coordinates explicitly exhibited.
The fermionic case is completely analogous and will not be treated; 
suffice it to mention that a fermionic extension is always the negative
of the corresponding bosonic extension.

We posit four types of oscillators $a_\al(\xx), a^\dagger_\al(\xx),
b_\al(\xx), b^\dagger_\al(\xx)$, subject to the nonzero commutators
\bes
[a_\al(\xx), b^\dagger_\bt(\yy)] &=& c_\ab\dlt(\xx-\yy), \nle
[a^\dagger_\al(\xx), b_\bt(\yy)] &=& - c^*_\ab\dlt(\xx-\yy). 
\eens
Define the real-valued field operators
\bes
\fa(\xx) &=& {1\/\sqrt2}(a_\al(\xx) + a^\dagger_\al(\xx)), 
\nlb{fapa}
\pa(\xx) &=& -{i\/\sqrt2}(b_\al(\xx) - b^\dagger_\al(\xx)),
\eens
which satisfy the CCR
\bes
[\fa(\xx), \pb(\yy)] &=& i\dlt_\ab\dlt(\xx-\yy).
\ee
From the oscillators we can also construct two more linearly independent
combinations 
\bes
\psi_\al(\xx) &=& {i\/\sqrt2}(a_\al(\xx) - a^\dagger_\al(\xx)), 
\nlb{psacha}
\chi_\al(\xx) &=& {1\/\sqrt2}(b_\al(\xx) + b^\dagger_\al(\xx)).
\eens
The normal-ordered combinations
\bes
E_\ab(\xx,\yy) &=& i\no{\fa(\xx)\pb(\yy)} \\
&=& \half(a_\al(\xx) b_\bt(\yy) + a^\dagger_\al(\xx) b_\bt(\yy) 
- b^\dagger_\bt(\yy) a_\al(\xx) - a^\dagger_\al(\xx) b^\dagger_\bt(\yy)),
\eens
satisfy the following central extension of $\gl(\infty)$:
\bes
[E_\ab(\xx,\xx'), E_\cd(\yy,\yy')] 
&=& \dlt_\cb E_\ad(\xx,\yy') \dlt(\yy-\xx') 
- \dlt_\ad E_\cb(\yy,\xx') \dlt(\xx-\yy') \nl
&+&\! \quart(c^*_\ad c_\cb - c^*_\cb c_\ad)\dlt(\xx-\yy')\dlt(\yy-\xx').
\ees
In particular, we know from (\ref{Jphi}) that what enters into the 
current algebra are the operators $E_\ab(\xx,\xx)$, where both space
indices are the same. 
\bes
[E_\ab(\xx,\xx), E_\cd(\yy,\yy)] 
&=& \dlt_\cb E_\ad(\xx,\xx) \dlt(\yy-\xx) 
- \dlt_\ad E_\cb(\xx,\xx) \dlt(\xx-\yy) \nl
&+&\! \quart(c^*_\ad c_\cb - c^*_\cb c_\ad)\dlt(\xx-\yy)\dlt(\zero).
\ees
If the matrices $M^a_\ab$ define a representation of the finite-dimensional
Lie algebra $\g$ as in (\ref{MM}), the smeared generators
\be
\J_X = -\int \dxx\, X^a(\xx) M^a_\ab E_\ab(\xx,\xx)
\label{smeared}
\ee
satisfy an extension of the current algebra $\map(d,\g)$ (\ref{mapg}):
\be
[\J_X, \J_Y] = \J_{[X,Y]} + \ext(X,Y),
\label{JXJY}
\ee
where $[X,Y]^a = if^{abc} X^b Y^c$ and the extension is
\bes
&&\ext(X,Y) 
= \quart \dlt(\zero) \int  \tr(XcYc^\dagger - Xc^\dagger Y c) 
\label{chirext}\\
&&=\ \quart \iint \dxx\,\dyy\, X^a(\xx) Y^b(\yy) 
M^a_{\bt\al} M^b_{\dlt\gm}
(c^*_\ad c_\cb - c^*_\cb c_\ad) \dlt(\zero) \dlt(\xx-\yy).
\eens
We see that the extension vanishes provided that 
$\tr(XcYc^\dagger) = \tr(Xc^\dagger Y c)$ for all pairs $X$ and $Y$.
If we write $c_\ab = \dlt_\ab + i\gm_\ab$ as in (\ref{cc}), the
condition for anomaly cancellation is equivalent to demanding that
$\tr(X\gm Y\dlt) = \tr(X\dlt Y\gm)$. If this condition does not hold,
the current algebra (\ref{JXJY}) is anomalous with a cycycle proportional
to $\dlt(\zero)$. Since there is no way to make sense of an infinite 
cocycle, the anomaly is inconsistent.

Another way to see that the extension (\ref{chirext}) is inconsistent
is to note that it is neither of the affine form (\ref{KMd}) nor of
the MF form (\ref{MF}). One can verify that if we replace $\dlt(\zero)$
by a finite number, the cocycle (\ref{chirext}) is not compatible with
the Jacobi identities.

An algebra that acts on the fields (\ref{fapa}) but not on (\ref{psacha})
is thus a ``chiral'', in our sense of the word: a chiral symmetry only acts
on half of the fields. The conclusion is that chiral theories have
serious problems with inconsistent anomalies. Although the analysis has
been carried out for bosonic fields, the same conclusion holds for 
fermions as well.

\section{ Point-splitting in time }
\label{sec:split}

We observed in section \ref{sec:ext} that spatial gauge algebras 
are not anomalous. To see the gauge anomalies of QJT, we must therefore 
consider time-dependent gauge transformations. This is not so
natural in the canonical formalism which deals with fields at fixed 
time. However, gauge anomalies do arise if we move away infinitesimally 
from the equal-time surface.

Recall from \cite{Lar08a} that absolute and relative fields are related by
\be
\phi_R(t,\xx) = \phi_A(t, \xx+\qq(t)), \qquad
\phi_A(t,\xx) = \phi_R(t, \xx-\qq(t)).
\label{fRA}
\ee
Fields with operator-valued arguments can be unabigously defined by
their Taylor series:
\bes
\phi_A(t,\xx) &=& \sum_\mm {1\/\mm!} \phi_{,\mm}(t)(\xx-\qq(t))^\mm, 
\nlb{spatjet}
\phi_R(t,\xx) &=& \sum_\mm {1\/\mm!} \phi_{,\mm}(t)\xx^\mm.
\eens
We employ multi-index notation which is explained e.g. in \cite{Lar08a}.
Since a $p$-jet is essentially the same thing as Taylor series
truncated at order $p$, this motivates the name QJT (Quantum Jet Theory).

Let us think about the physical meaning of normal ordering. An 
annihilation operator destroys a particle at time $t$, and a creation
operator recreates it an instant later. Denote the duration of this 
instant by $2\eps$, so that annihilation takes place at time $t-\eps$
and creation at time $t+\eps$, but both take place at
the same absolute location $\xx$. Although the distance to the fixed
origin remains the same, the distance from the 
observer changes; if the observer moves at constant velocity  $\uu$, 
i.e. $\qq(t) = \uu t$, the relative location changes from 
$\xx + \uu\eps$ to $\xx - \uu\eps$. 
The relation between absolute and relative fields yields
\bes
\phi_A(t+\eps,\xx) &=& \phi_R(t+\eps, \xx-\uu \eps), \nle
\phi_A(t-\eps,\xx) &=& \phi_R(t-\eps, \xx+\uu \eps)
\eens
In QJT we deal with relative rather than absolute fields, and hence
we consider the relative creation and annihilation operators
\bes
a_\al(\xx) &=& \fa(\xx+\eps\uu) 
\approx \fa(\xx) + \eps\d_\uu\fa(\xx), \nl
a^\dagger_\al(\xx) &=& \fa(\xx-\eps\uu) 
\approx \fa(\xx) - \eps\d_\uu\fa(\xx), \nle
b_\al(\xx) &=& -i\pa(\xx+\eps\uu) 
\approx -i\pa(\xx) - i\eps\d_\uu\pa(\xx), \nl
b^\dagger_\al(\xx) &=& -i\pa(\xx-\eps\uu) 
\approx -i\pa(\xx) + i\eps\d_\uu\pa(\xx).
\eens
where $\d_\uu = u_i \d_i$ is the directional derivative in the direction
of the observer's velocity. Here and henceforth we suppress the subscript 
``R'' for relative, and equip the fields with a discrete index $\al$.
{F}rom the canonical brackets
\be
[\fa(\xx), \pb(\yy)] = i\dlt_\ab \dlt(\xx-\yy),
\label{fapb}
\ee 
we read off the following nonzero brackets, to lowest order in $\eps$:
\bes
{[}a_\al(\xx), b^\dagger_\bt(\yy)] &=& 
\dlt_\ab(\dlt(\xx-\yy) + 2\eps\d_\uu\dlt(\xx-\yy)), \nl
{[}a^\dagger_\al(\xx), b_\bt(\yy)] &=& 
\dlt_\ab(\dlt(\xx-\yy) - 2\eps\d_\uu\dlt(\xx-\yy)), \nle
{[}a_\al(\xx), b_\bt(\yy)] &=& \dlt_\ab\dlt(\xx-\yy), \nl
{[}a^\dagger_\al(\xx), b^\dagger_\bt(\yy)] &=& \dlt_\ab\dlt(\xx-\yy).
\eens
This is reminiscent in (\ref{chirality}), where we also have four types 
of oscillators. The situation is not completely identical, since the
$[a,b]$ and $[a^\dagger,b^\dagger]$ brackets do not vanish, but what
matters is that the combinations
\bes
\fa(\xx) &=& \half(a_\al(\xx) + a^\dagger_\al(\xx)), \nle
\pa(\xx) &=& {i\/2}(b_\al(\xx) + b^\dagger_\al(\xx)),
\eens
satify (\ref{fapb}) to lowest order in $\eps$. We define the
normal-ordered bilinears
\bes
E_\ab(\xx,\yy) &=& i\no{\fa(\xx)\pb(\yy)} \nl
&=& -\quart\no{ (a_\al(\xx) + a^\dagger_\al(\xx))
(b_\bt(\yy) + b^\dagger_\bt(\yy)) } \\
&=& -\quart(a_\al(\xx)b_\bt(\yy) + a^\dagger_\al(\xx)b_\bt(\yy)
 + b^\dagger_\bt(\yy)a_\al(\xx) + a^\dagger_\al(\xx) b^\dagger_\bt(\yy) ).
\eens
Equivalently, normal ordering amounts to the redefinition
\bes
E_\ab(\xx,\yy) &\to&  E_\ab(\xx,\yy) 
- \quart [b^\dagger_\bt(\yy), a_\al(\xx)] 
\\
&=& E_\ab(\xx,\yy) 
+ \quart \dlt_\ab( \dlt(\xx-\yy) - 2\eps\d_\uu\dlt(\xx-\yy) ).
\eens
The normal-ordered generators satify the following central extension of
$\gl(\infty)$:
\bes
&&[E_\ab(\xx,\xx'), E_\cd(\yy,\yy')] = 
\label{noEE} \\
&&\qquad =\ \dlt_\cb E_\ad(\xx,\yy') \dlt(\yy-\xx') 
- \dlt_\ad E_\cb(\yy,\xx') \dlt(\xx-\yy') \nl
&&\qquad\qquad +\ \quart \eps \dlt_\ad \dlt_\cb 
(\dlt(\yy-\xx')\d_\uu\dlt(\xx-\yy') - \dlt(\xx-\yy')\d_\uu\dlt(\yy-\xx') ).
\eens
The smeared $\map(d,\g)$ generators (\ref{smeared}) therefore satisfy
an extension of the current algebra (\ref{JXJY}), with
\bes
\ext(X,Y) &=&
\quart \eps \iint \dxx\,\dyy\, \tr(X(\xx)Y(\yy)) \times \nl
&&\qquad\times\, 
(\dlt(\yy-\xx)\d_\uu\dlt(\xx-\yy) - \dlt(\xx-\yy)\d_\uu\dlt(\yy-\xx) ) \nl
&=& \half \eps \dlt(\zero) \int \dxx\, \tr(X(\xx) \d_\uu Y(\xx)).
\ees
If we define $K = \eps\dlt(\zero)/2$, the algebra becomes
\be
[\J_X, \J_Y] = \J_{[X,Y]} + K \int \dxx\, \tr(X(\xx) \d_\uu Y(\xx)).
\label{JXJYK}
\ee
This is equivalent the central extension $\Aff(d,\g)$ in the form 
(\ref{KMd1}), and the relation between $\J_X$ and $J^a(t,\xx)$ is
\be
\J_X = \int \dxx\, X^a(\xx-\uu t) J^a(t,\xx).
\ee

The extension (\ref{JXJYK}) is the main result of this paper. It shows
that although the gauge anomalies of QJT do not appear on a fixed 
foliation, they do show up when we work with relative fields, provided
that we move away an infinitesimal distance from the equal-time surface.
Note that there is no assumption about chirality here; the central 
extension appears for all gauge groups and all non-trivial irreps. 
The extension can of course be made to cancel by matching bosonic and
fermionic degrees of freedom.

The derivation above is certainly formal. The constant $K$ is a 
product of the form $0\times\infty$, which can be anything. To give a
definite value to this expression, we must consider a regularized
theory, where the time-split $\eps$ is small but finite, and the
delta-function $\delta(\zero)$ stands for large but finite number.
Their product $K$ is then well defined, and by choosing $\eps$ suitably,
$K$ can be given a finite but nonzero limit. This process is
reminiscent of renormalization, and therefore one may expect that
similar anomalies may arise if we consider point-splitting in a
renormalizable field theory, formulated in terms of relative fields.

The method in this section does not generalize to
general-covariant theories. This is not surprising, since the
point-splitting prescription leads to the central extension of
$\gl(\infty)$ (\ref{noEE}). Since the Virasoro extension of $\vect(d+1)$
is not central, it can not possibly arise in this way. The underlying
assumption that the fields are of the form (\ref{fRA}) is not valid,
because a spacetime diffeomorphism will in general modify the foliation
of spacetime into space and time. The foliation is only preserved by
spatial diffeomorphisms, but as we saw in section \ref{sec:ext}, the
spatial algebra $\vect(d)$ is anomaly free. The relevant $\Vir(d+1)$
extensions do arise in the manifestly covariant formalism described in
subsection \ref{ssec:MCCQ} below, where a foliation is avoided
altogether.

\section{Quantum Jet Theory}

\subsection{ Manifestly covariant canonical quantization}
\label{ssec:MCCQ}

The decomposition of spacetime into space and time is a drawback of
canonical quantization. This problem becomes particularly serious when
one wants to study the constraint algebra of a background-independent
theory like general relativity, since four-diffeomorphisms do not
preserve the foliation. However, as was noted already by Lagrange, the
notion of phase space is itself covariant; it is the space of histories
which solve the equations of motion. Such a history can be coordinatized
by the values of the position and velocity, or momentum, at time $t=0$,
but this is only one way to put coordinates on phase space.

The idea behind Manifestly Covariant Canonical Quantization (MCCQ),
introduced in \cite{Lar04,Lar06a}, was to quantize in the history phase
space first, and then impose dynamics in a BRST-like manner afterwards.
It is very similar to BV quantization as described in \cite{HT92},
except that there are twice as many degrees of freedom; in addition to
the fields and antifields, we also introduce the corresponding momenta.
This is necessary because in order to do canonical quantization we need
an honest Poisson bracket, and in conventional BV quantization there is
only an antibracket.

Let us return to the abbreviated notation where the index $\al$ stands
for both discrete indices and spacetime coordinates, and contraction
includes both contraction of discrete indices and integration over
spacetime. Let us consider some theory with spacetime fields, or
histories, $\phi^\al$ and action $S[\phi]$. The covariant phase space is
the space of histories $\phi^\al$ which solve the Euler-Lagrange
equations
\be
\EE_\al = {\dlt S\/\dlt \phi^\al} = 0.
\label{Ea}
\ee
Unfortunately, this description of phase space is not very convenient,
because it requires us to find the general solution to the Euler-Lagrange
equations. Fortunately, we are not interested in phase space itself, but
rather in the dual space of functions over it; this is what becomes our 
Hilbert space after quantization. This space has a nice cohomological
description. For each Euler-Lagrange equation $\EE_\al$,
introduce an antifield $\fsa$ of opposite Grassmann parity. For
simplicity, we assume that $\phi^\al$ and thus $\EE_\al$ are bosonic,
which means that $\fsa$ is a fermionic field. Furthermore, we introduce
momenta both for the field and for the antifield, subject to the nonzero
graded commutators
\be
[\phi^\al, \pb] = i\dlt^\al_\bt, \qquad 
\{\fsa, \psb\} = \dlt_\al^\bt
\ee
We can now impose dynamics by passing to the zeroth cohomology group of 
the fermionic BV operator
\be
Q = \EE_\al \psb.
\ee
Indeed, we have
\be
[Q, \phi^\al] = 0, \qquad \{Q, \fsa\} = \EE_\al.
\ee
Hence all antifields disappear from the cohomology because they are not
closed, whereas the fields which do not solve the Euler-Lagrange equation
vanish because they are not exact.

Alas, this construction does not quite work even for the harmonic 
oscillator. The problem is that spurious cohomology arises due to an
overcounting of the degrees of freedom. To eliminate this overcounting, 
one must first identify momenta and velocities, and also compensate for
the existence of solutions; the second type of problem already arises in
conventional BV quantization. These problems can be solved by introducing
further, second-order antifields \cite{Lar07}, at least for the harmonic 
oscillator, but the extra antifields are noncovariant and ugly, and the
formalism becomes very complicated.

Despite these problems, the original formalism has great appeal when it
comes to gauge theories, because the constraint algebras naturally act
in spacetime. In particular, the spacetime form of the constraint algebra
for general relativity is the four-diffeomorphism algebra $\vect(4)$. In
non-covariant canonical quantization, we must respect the foliation of 
spacetime, which modifies $\vect(4)$ into the Dirac algebra (however,
see \cite{KR95,Mar96}). In contrast, in a manifestly covariant formalism,
the constraint algebra is $\vect(4)$ itself, which means that the
representation theory of $\Vir(4)$ applies.

Moreover, the notion of quantization is very natural in the history 
space; it simply amounts to introducing a vacuum that is annihilated by 
all negative frequency modes. When we introduce oscillators for the free
electromagnetic field in (\ref{aaEM}), we associated $a^\dagger_i(\kk)$ 
with positive energy and  $a_i(\kk)$ with negative energy, and demanded
that the negative energy modes annihilate the vacuum: $a_i(\kk)\ket0 = 0$.
The treatment in history space is analogous. Each spacetime history
$\phi^\al(t,\xx)$ can be Fourier transformed in the time coordinate, and the
Fourier modes $\phi^\al(k_0,\xx)$ with $k_0 < 0$ are defined to annihilate the
vacuum. Therefore we need to construct representations of the constraint
algebra, be it $\map(4,\g)$ or $\vect(4)$, that are of lowest-energy type.

\subsection{ The need for QJT }

To build lowest-energy representations of $\vect(1)$ is easy. The recipe
consists of the following steps:
\begin{itemize}
\item
Start from a classical module, i.e. a scalar density a.k.a. a primary
field.
\item
Introduce canonical momenta.
\item
Normal order.
\end{itemize}
As is well known, this recipe results in the central extension known as
the Virasoro algebra (\ref{Vir}).

There are two main obstructions which prevent a straightforward 
generalization of this technique from $d=1$ to higher dimensions:
\begin{itemize}
\item
Fields in $d>1$ dimensions do not admit a unique polarization, 
analogous to the division into positive and negative frequency
modes on the circle.
\item
Normal ordering does not suffice to eliminate all infinities; some 
kind of further renormalization is necessary.
\end{itemize}
The insight in the seminal paper \cite{RM94}, geometrically clarified
in \cite{Lar98}, is that one should not start from a classical
representation on tensor fields. Instead, one must pass to the 
corresponding space of $p$-jets, and consider one-dimensional 
histories in this space. The crucial point is that a $p$-jet
history consists of {\em finitely} many functions of a {\em single}
variable, and therefore the two problems above do not arise. However,
since the classical $\vect(d)$ realization on $p$-jets is nonlinear,
the resulting extension (\ref{mVir}) is non-central; the $S^\mu(k)$
do not commute with diffeomorphisms except when $d=1$.

The $\Vir(d+1)$ realization in QJT is naturally expressed in terms of
covariant spacetime $p$-jet histories rather than non-covariant 
spatial jets (\ref{spatjet}); e.g., 
\be
\phi_A(x,t) &=& \sum_m  {1\/m!} \phi_{,m}(t)(x-q(t))^m.
\label{covjet}
\ee
Here $m = (m_0, m_1, ..., m_d) \in \ZZ^{d+1}$ is a $(d+1)$-dimensional
multi-index and $t$ is a timelike parameter; spatial jets are recovered
by demanding that $x^0 = q^0(t) = t$, which makes the Taylor series
(\ref{covjet}) independent of $m_0$. Covariant jets are spanned by the 
operator-valued functions $q^\mu(t), \phi_{,m}(t)$, which
together with their canonical momenta $p_\mu(t), \pi^{,m}(t)$ satisfy the 
Heisenberg algebra
\bes
[q^\mu(t), p_\nu(t')] &=& i\dlt^\mu_\nu \dlt(t-t'), 
\nlb{histccr}
{[}\phi_{,m}(t), \pi^{,n}(t')] &=& i \dlt^n_m \dlt(t-t').
\eens
In contrast to the spatial jets (\ref{spatjet}), these commutation 
relations live in the history phase space, so field operators at 
different values of $t$ commute.
The $\vect(d+1)$ generators can be written as
\bes
L_\mu(k) &=& \int dt\ \bigg( \exp(ik\cdot q(t)) p_\mu(t) + 
\nlb{Lqjt}
&&+ \sum_{0\leq |n| \leq |m| \leq p} \pi^{,m}(t) 
T^n_m( \exp(ik\cdot q(t))\d_\mu) \phi_{,n}(t) \bigg),
\eens
where we define $T^m_n(\xi)$ for every vector field 
$\xi = \xi^\mu\d_\mu$ by
\bes
T^m_n(\xi) &=& \sum_{\mu,\nu = 0}^d {n\choose m} 
\d_{n-m+\nu}\xi^\mu T^\nu_\mu +
\label{Tmn}\\
&&\  +\ \sum_{\mu = 0}^d {n\choose m-\mu}\d_{n-m+\mu}\xi^\mu 
\ -\ \sum_{\mu = 0}^d \dlt^{m-\mu}_n \xi^\mu,
\eens
and the matrices $T^\mu_\nu$ satisfy $\gl(d)$:
\be
[T^\mu_\nu, T^\rho_\si] = 
\dlt^\mu_\si T^\rho_\nu - \dlt^\rho_\nu T^\mu_\si.
\ee
The integrand in (\ref{Lqjt}) only depends on functions of a single
variable $t$, and it can therefore be Fourier transformed. Normal 
ordering with respect to the corresponding frequency then yields a
realization of the multi-dimensional Virasoro algebra (\ref{mVir}).
Hence we obtain a new well-defined representation of $\Vir(d+1)$ for each
$\gl(d+1)$ representation $\varrho$ and for each jet order $p$. 

Analogously, a $\map(d+1,\g)$ representation is given by
\bes
J^a(k) &=& \sum_{0\leq |n| \leq |m| \leq p}
{m \choose n} i^{|m-n|} k^{m-n} \times \nle
&&\times\, \int dt\, \exp(ik\cdot q(t)) \pi^{,m}(t) M^a \phi_{,n}(t),
\eens
where $k^m = k_0^{m_0} k_1^{m_1} ...k_d^{m_d}$ and
the matrices $M^a$ satisfy $\g$. After normal ordering w.r.t. frequency,
this expression yields a representation of $\Aff(d+1)$ (\ref{KMd}).

\subsection{ The need for gauge anomalies }

According to conventional wisdom, gauge anomalies are a sign of
inconsistency and must necessarily cancel. This viewpoint is natural if
we think of gauge symmetries as redundancies of the description, but
it is an oversimplified point of view. What is important is not whether
a gauge symmetry is represented trivially or not, but whether the
resulting quantum theory is unitary.  It is quite possible that the
Hilbert space of an anomalous gauge theory has a positive-definite inner
product and is thus consistent; the subcritical free string is a
well-known example \cite{GSW87}. It is of course not consistent
to try to eliminate a gauge symmetry in the presence of anomalies; a
gauge anomaly transforms a classical gauge symmetry into a quantum
global symmetry, which acts on the Hilbert space rather than reducing
it. The resulting theory may (subcritical free string) or may not
(anomalous chiral fermions, supercritical free string) be consistent.

Gauge anomalies may not only be consistent but even necessary.
It was pointed out in \cite{Lar06b} that local gauge symmetries must act
nontrivially the presence of a nonzero charge, provided that we consider
the natural completion of the gauge algebra which also contains 
generators that diverge at infinity. This completion inevitably arises
if we expand the gauge symmetry in a Laurent series. E.g., consider
Yang-Mills theory in three dimensions. The $\map(3,\g)$ Laurent modes,
\be
J^a_{n,\ell,m} = r^n Y_{\ell,m}(\theta,\varphi) M^a,
\ee
where the matrices $M^a$ define a finite-dimensional representation of 
$\g$, can be classified according to their behavior at $r = \infty$:
local ($n<0$), global ($n=0$), and divergent ($n>0$).
Since
\be
[J^a_{n,0,0}, J^b_{-n,0,0}] = if^{abc} J^c_{0,0,0},
\ee
we cannot assume that the local generator $J^b_{-n,0,0}$ is represented
trivially if the global charge operator $J^c_{0,0,0}$ is nonzero. 

At this point, unitarity may seem to require a gauge anomaly, 
in the same way as nontrivial representations of affine algebras are only
possible with a nonzero central extension. However, this only follows if
we assume that the representations are of lowest-weight type. The
spatial subalgebra is not expected to be of this type; after all, it
is energy that is bounded from below, not the spatial components of
energy-momentum. This is consistent with the observation in (\ref{spat})
that the spatial subalgebra $\vect(d)\ltimes\map(d,\g)$ is anomaly
free. In contrast, the relevant representations of the spacetime 
algebras are of lowest-energy type, and hence we expect that gauge 
anomalies arise, of the form $\Vir(d+1)\ltimes\Aff(d+1,\g)$.

But even in the absense of gauge anomalies, the argument above shows
that nonzero charge and divergent transformations inevitably lead to
a nontrivial representation of the local gauge transformations. This 
argument relies on the passage to the natural completion of
the gauge algebra. By doing so, we must enlarge the Hilbert space to
encompass new states obtained by acting with divergent operators. 
These states are not gauge invariant, only gauge covariant. The local
operators with $n<0$ do of course still annihilate the original
Hilbert space. Hence there is no contradiction between gauge invariance
of the original Hilbert space of physical states, and anomalous gauge
covariance of the completed Hilbert space which also carries an action
of the divergent generators.

\subsection{ Finite anomalies and four dimensions}

Although gauge and diff anomalies of the right kind are
desirable, they must be finite; infinite anomalies are a disaster. QJT
suggests a natural regularization, where we work with $p$-jets rather
then infinite jets, i.e. the Taylor series (\ref{spatjet}) or
(\ref{covjet}) are truncated at order $p$. The abelian charges, i.e. the
values of the parameters multiplying the cocycles, depend on $p$. They
are always finite in the regularized theory, but generically
diverge in the field theory limit $p\to\infty$. Taken at face value, 
this appears to be a serious problem for QJT.

However, it was noted in \cite{Lar04,Lar06a,Lar07} that the divergent
parts of the anomalies can be made to cancel between bosonic and
fermionic degrees of freedom, leaving a finite but nonzero remainder,
but only if spacetime has four dimensions.
This result depends only on a few natural assumptions: the model has
both bosonic and fermionic fields, the equations of motion are first
order for fermions and second order for bosons, there are bosonic gauge
symmetries leading to third order continuity equations, and there are
no reducible gauge symmetries. The gauge symmetries can then only be
implemented up to order $p-3$, and require ghosts of order $p-2$. Under
these assumptions, the algebras of spacetime diffeomorphisms, gauge
transformations, and reparametrizations of the observer's trajectory can
all be made to acquire finite but nonzero anomalies in exactly four
spacetime dimensions. 

Let us illustrate the counting for electrodynamics. There are some
fermionic fields $\psi$ which satisfy the first-order Dirac equation
$\gamma^\mu(i\d_\mu + eA_\mu)\psi = 0$. The bosonic field is the
gauge potential $A_\mu$, with second-order equation of motion
$\d_\nu F^\mn = j^\mu$, and the continuity equation comes from the
third-order (in $A$) identity $\d_\mu\d_\nu F^\mn \equiv 0$. 

Unfortunately, on closer scrutiny the counting becomes less appealing,
because it appears to predict a field content in disagreement with
experimental data. However, the situation remains unclear, and I remain
optimistic that the problems can be circumvented once QJT is better
understood. The prediction of four spacetime dimensions is quite robust,
because it makes it possible to cancel the infinite parts of no less
than five different anomalies in QJT.

\section{ Conclusion }

The existence of new diff and gauge anomalies proves that QJT is
substantially different from QFT. Given that QFT is incompatible with
gravity, this is very positive. As is well known from conformal field
theory, regarded as diffeomorphism-symmetric field theory on the circle,
Virasoro-like anomalies are necessary to combine diffeomorphism symmetry
with locality, in the sense of correlators depending on separation.
A non-holographic quantum theory of gravity with local
observables hence requires diff anomalies and QJT.

\end{document}